\documentstyle[12pt]{article}
\begin{document}
\begin{center}
{\large\bf UN/ESA WORKSHOPS ON BASIC SPACE SCIENCE:\\ 
AN INITIATIVE IN THE WORLDWIDE DEVELOPMENT OF ASTRONOMY}\\[0.5cm]
{\bf Hans J. Haubold}\\
Office for Outer Space Affairs\\
United Nations\\
Vienna International Centre\\
A-1400 Vienna, AUSTRIA\\
E-mail: haubold@kph.tuwien.ac.at\par
\end{center}
\noindent
{\bf ABSTRACT}\par
\medskip
\noindent
In 1990, the United Nations, in cooperation with the European Space Agency, initiated the organization of a series of annual Workshops on Basic Space Science for the benefit of astronomers and space scientists in Asia and the Pacific, Latin American and the Caribbean, Africa, Western Asia, and Europe. This article summarizes accomplishments of these Workshops (1991-1998) and their follow-up projects with a view to enhance the worldwide development of astronomy and space science. The Workshops are being considered unique and a model for such an endeavor.\par
\par
\bigskip
\noindent 
{\large\bf 1 OVERVIEW}\par
\medskip
\noindent
The last decade of the 20th century is using telescopes in space, in aircraft, on the ground, and even underground to address fundamental questions concerning our place in the Universe. Do planets orbit nearby stars?, What triggers the formation of stars?, How do life-giving elements such as carbon and oxygen form and disperse throughout the galaxy?, Where can black holes be found, and do they power luminous galaxies and quasars?, How and when did galaxies form?, Will the Universe continue to expand forever, or will it reverse its course and collapse on itself?, What is the physical nature of dark matter?. These questions, the answers (some of them of tentative nature), the instruments devised to find them, and the observations and recommendations, both on the national level of the United States and in Canada, Japan, and European countries, are explored in a document titled The Decade of Discovery in Astronomy and Astrophysics and subsequent publications. The great achievements made in the development of planetary explorations, astronomy, and astrophysics in the 90s can be comprehensively assessed by looking through the pages of these documents (Bahcall, 1991; Brown, 1990). 

     Astronomy has traditionally been an international enterprise: Because astronomy has deep roots in virtually every human culture; because it helps us to understand our place in the vast scale of the Universe; and because it teaches us about our origins and evolution. By virtue of these facts the astronomical community has long shown leadership in creating international collaborations and cooperation. The International Astronomical Union (IAU) was the first of the modern international scientific unions organized under the Versailles treaty. The names of the European Southern
Observatory, the International Ultraviolet Explorer, or World Space Observatory are self-explanatory. Despite all of this, the level reached in practicing and teaching of astronomy is surprisingly unequal across countries around the world. 

     Of the 185 Member States of the United Nations (UN), nearly 100 have professional or amateur astronomical organizations. In only about 60 of these countries, however, has professional astronomical research and education been developed to such a level that they have joined the International Astronomical Union as members. And only about 20 countries, representing 15\% of the world's population, have access to the full range of modern, front-line astronomical facilities and information. This does not include most of the Eastern European, Baltic, and former countries of the Soviet Union, whose fragile economies keep them from achieving their full potential, despite the excellence of their astronomical heritage and education (Andersen, 1998).

     This paper provides details of an initiative lead by the United Nations and the European Space Agency (ESA) in the development of astronomy on a worldwide basis, thereby focusing on developing nations.\par
\medskip
\noindent
{\bf 1.1 UN/ESA Workshops on Basic Space Science for Developing Countries: Asia and the Pacific, Latin America and the Caribbean, Africa, Western Asia}\par
\smallskip
\noindent
In 1991, the United Nations, in cooperation with the European Space Agency$^1$, held its first Workshop on Basic Space Science (hereafter shortly named UN/ESA Workshops) in India for Asia and the Pacific region (Haubold and Khanna, 1992; UN GA doc of 17 September 1991). Since then, such Workshops have been held annually in the different regions around the world to make a unique contribution to the worldwide development of planetary exploration and astronomy, particularly in developing countries (Haubold et al., 1995; Haubold, 1996; Haubold and Wamsteker, 1997) Workshops were held in 1992 in Costa Rica and Colombia for Latin America and the Caribbean (Fernandez and Haubold, 1993; Haubold and Torres, 1994; UN GA doc of 20 January 1993), in 1993 in Nigeria for Africa (Haubold and Onuora, 1994; UN GA doc of 26 May 1994), and in 1994 in Egypt for Western Asia (Haubold and Mikhail, 1995a, b; UN GA doc of 11 August 1994). A month-to-month update on results and new developments related to the UN/ESA Workshops is made available on the Workshop's homepage$^2$. For all these Workshops an unanimously adopted formula was established, saying that the host country covers a major part of the internal expenses while UN and ESA provide financial resources for international air travel for up to 25 astronomers and space scientists from developing nations. Co-organizing entities1 and many other institutions and universities around the world paid for expenses of distinguished scientists from industrialized countries. The total number of Workshop participants was limited to around 75 to secure intense and lively interaction among all individuals participating at and contributing to the Workshops. To emphasize a high commitment from all parties concerned, the Governments of the host countries of the Workshops entered into a highly tuned Memorandum of Understanding with the United Nations with respect to legal and logistic matters relevant to the Workshops. Needless to say that astronomers and space scientists from any country of the regions were able to attend the Workshops.

     The selection of astronomical topics for the scientific programme of the Workshops depended on the interest of the local organizers taking into account the regional level of the development of the field. A vital part of the programme of each Workshop was the late afternoon working group sessions which provided all participants with a common platform to make constructive observations and recommendations addressing the status of astronomy in their specific region. Region by region these observations and recommendations have been collected and published as part of UN documents and Proceedings of the Workshops (Haubold and Khanna, 1992; UN GA doc of 17 December 1991; Fernandez and Haubold, 1993; Haubold and Torres, 1994; UN GA doc of 20 January 1993; Haubold and Onuora, 1994; UN GA doc of 26 May 1994; Haubold and Mikhail, 1995a, b; UN GA doc of 11 August 1994; UN GA doc of 14 May 1996; UN GA doc of 20 January 1998; Haubold and Mezger, 1998; UN GA doc of 13 December 1996a, b).\par
\medskip
\noindent 
{\bf 1.2 Inauguration of the Astronomical Telescope Facilities in Sri Lanka and Honduras}\par
\smallskip
\noindent
Interest in the design and construction of very large space-based and ground-based telescopes has never been greater than at the present time and there is no shortage of projects for such instruments. The excitement and opportunities associated with the very large is limitless but can easily obscure the fact that small telescopes have an increasingly important place in modern astronomical research. It is vital that this assertion be understood and widely accepted: There are many institutions, even entire countries, who cannot afford to enter the race for giant telescopes and there are astronomical observational programmes better pursued with small telescopes. There is a need for more small telescopes, for better use of existing ones, their networking, and for continual upgrading of the instrumentation used on them (Warner, 1986).

     For 1995, the UN/ESA Workshop was held in Sri Lanka to inaugurate an astronomical telescope facility, following the donation of an astronomical telescope from the Government of Japan to the Government of Sri Lanka, at the Arthur C. Clarke Centre for Modern Technologies$^3$ ( Haubold, 1999; UN GA doc of 14 May 1996). In 1997, the UN/ESA Workshop inaugurated the Central American Astronomical Observatory at Tegucigalpa$^4$, Honduras, with the dedication of the Telescopio Rene Sagastume Castillo at the Suyapa Observatory for six Central American countries (Costa Rica, El Salvador, Guatemala, Honduras, Nicaragua, Panama) (Haubold, 1999; UN GA doc of 20 January 1998).\par
\medskip
\noindent 
{\bf 1.3 Europe: Assessment of the Achievements of the Workshops}\par
\smallskip
\noindent
Based on a request from the United Nations, the Foreign Office of the Government of Germany, through the German Space Agency (DARA), made it possible to hold a UN/ESA Workshop at the Max-Planck-Institute for Radioastronomy (MPIfR), Bonn, Germany, in 1996, for the benefit of Europe. This Workshop analyzed the results of all previous UN/ESA Workshops, particularly the follow-up projects that emanated from the Workshops (which are reviewed in sections 2 and 3 of this paper), and charted the course to be followed in the future. Additional to this objective, taking advantage of the host country's sophisticated high level of research in astronomy and astrophysics, the Workshop addressed scientific topics at the forefront of research in such diverse fields as photon, neutrino, gravitational wave, and cosmic ray astronomy, respectively (Haubold and Mezger, 1998; UN GA doc of 13 December 1996a). Because Europe does not comprise developing countries, a special appeal to a number of European countries had to be initiated, leading to the generous concurrence of Germany to host a Workshop of this nature. Not all individuals concerned agreed with such a decision (Elsaesser, 1995) while other praised the achievements of this Workshop (Kogure, 1997).\par
\medskip
\noindent
{\bf 1.4 A Case for Planetaria}\par
\smallskip
\noindent
The Universe is more than just a distant wonder; it gives to every human the gift of the starry nighttime sky. This gift is free to all, except when veiled by clouds or masked by human lights. It is an environmental treasure that can be shared by all people and hoarded by none. In an era when the light-saturated sky of cities has transformed the stars into an endangered species, the beauty and mystery of the night sky is preserved upon the domes of the world's planetaria. A planetarium is a theater that can preserve cultural heritage unique to every nation and retell the ancient stories beside the modern science. Planetaria are centres of science education for the public and the professional as well and have a broad impact on public education. Many young people are drawn to a career in science and technology by an early interest in astronomy and space, frequently facilitated trough activities of planetaria.

1992 had been designated as International Space Year (ISY) by a wide variety of national and international space organizations, including the United Nations. To help generate interest and support for planetaria as centres of culture and education, the United Nations in cooperation with the International Planetarium Society (IPS)$^5$, as part of its ISY activities, published a guidebook titled Planetarium: A Challenge for Educators$^6$ (Smith and Haubold, 1992). Later, this booklet was translated by national planetarium associations from English into Japanese, Slovakian, and Spanish, and is still available from the United Nations. Currently it is being translated into Arabic. Since the publication of this guidebook, that was requested and distributed worldwide in more than 4000 copies, the United Nations facilitates the establishment of planetaria (Kitamura, 1994; Vietnam, 1995).\par
\clearpage
\noindent
{\bf 1.5 Donation of ESA Computers to Astronomical Institutions in Developing Countries}\par
\smallskip
In 1993, the European Space Agency, through the United Nations, donated 30 personal computer systems for use at universities and research institutions in Cuba, Ghana, Honduras, Nigeria, Peru, and Sri Lanka (Haubold et al., 1995; Wamsteker, 1994).\par
\medskip
\noindent 
{\bf 1.6 The United Nations International Conference on Near-Earth Objects}\par
\smallskip
\noindent
The odds of our international society getting obliterated by an asteroid impact in the next 50 years has been estimated at 6000 to 1. That probability and its consequences, as well as the methods for discovery and mitigation, are described in a textbook by 120 authors in 46 chapters (Gehrels, 1994). The first step in mitigation is to discover with astronomical telescopes the estimated 1700 near-Earth objects that are larger than 1-km in diameter, the ones energetic enough to cause serious damage on a worldwide scale.

     In 1995, scientists from around the world gathered at the United Nations Headquarters in New York to discuss a broad range of scientific issues associated with near-Earth objects. This gathering became known as the first United Nations International Conference on Near-Earth Objects$^7$ (Haubold and Remo, 1998; Remo, 1997). Among the topics addressed by the Conference was the establishment of adequate observational facilities for near-Earth objects in both the northern and southern hemispheres. As a first step, these facilities should be improvements of existing astronomical telescopes, particularly small ones, including those in developing nations. Observation programmes could be coordinated with activities of amateur astronomy groups and organized on an international level which may lead to the establishment of a network of moderate-sized astronomical telescopes as discussed in the UN/ESA Workshops and supported by the European Spaceguard Foundation and the Japanese Spaceguard Association (UN GA doc of 14 May 1996; UN GA doc of 20 January 1998).\par
\bigskip
\noindent
{\large\bf 2 ASTRONOMY BENEFITS FROM FOLLOW-UP\\ 
PROJECTS OF THE UN/ESA WORKSHOPS}\par
\medskip
\noindent
Additional to the benefits of common scientific Workshops, the UN/ESA series has lead to the implementation of a number of follow-up projects in developing countries where the respective Governments and astronomical communities did actively participate in the organization of the Workshops (Haubold et al., 1995; Haubold, 1996;
Haubold and Wamsteker, 1997).\par
\medskip
\noindent
{\bf 2.1 Asia and the Pacific: Sri Lanka's Telescope Facility}\par
\smallskip
\noindent
The Arthur C. Clarke Centre for Modern Technology (ACCMT) was established in 1984 with the objective to accelerate the process of introduction and development of modern technologies in Sri Lanka in the fields of computers, communications, space technologies, robotics, and energy (Gehrels, 1984; Gehrels, 1988). In 1991, a team of scientists who represented Sri Lanka at the first UN/ESA Workshop at Bangalore, India, indicated the importance of acquiring an astronomical telescope for Sri Lanka. At this Workshop it was recommended and widely supported that three observatories should be established in Sri Lanka. Subsequently, the United Nations Office for Outer Space Affairs recommended to the Government of Japan to consider the donation of a telescope to Sri Lanka in order to implement the programme (Kitamura, 1992; Wickramasinghe, 1993). The Government of Japan favorably analyzed the request and offered a 45-cm Cassegrain reflecting telescope to the Government of Sri Lanka through the Cultural grant-in-aid scheme. A team of officials from the Government of Japan and the United Nations visited Sri Lanka in 1992 and a meeting was held at the Sri Lankan Association for the Advancement of Science in Colombo. A decision was taken to install this telescope at the Department of Meteorology in Colombo. Due to heavy expenditures involved with the physical infrastructure needed for the telescope and considering its technical and scientific capabilities, ACCMT was requested to take over the project. It was decided to install this telescope at ACCMT because a new four story building was being constructed at the time and the construction of the top floor could be modified to house the telescope facility and ACCMT has the capability to handle the repair and maintenance of the telescope including the fully automated electronic equipment. It was also decided to construct the telescope room with a sliding roof instead of a rotating doom due to the high costs involved. In 1994, the Board of Governors of ACCMT formed a Steering Committee comprised of astronomers, physicists, and engineers to implement the project. The Cabinet of Sri Lanka eventually approved the project and the 45-cm reflecting telescope from the GOTO Manufacturing Company of Japan arrived in Sri Lanka in the middle of 1995. The astronomical telescope facility was inaugurated during the fifth UN/ESA Workshop, hosted by the Government of Sri Lanka, in January 1996 at ACCMT. At present, the Space Applications Division of the Space Applications Centre of ACCMT is managing and operating the astronomical telescope facility (Haubold, 1999; UN GA doc of 14 May 1996).\par
\medskip
\noindent  
{\bf 2.2 South America: Colombia's Radio Telescope}\par
\smallskip
\noindent
On the occasion of holding the second UN/ESA Workshop at Bogota, Colombia, in 1992, the Galactic Emission Mapping (GEM) project was established with the aim of obtaining a full sky, accurate and multi-frequency survey of the galactic radio emission (408 to 5000 MHz) (Haubold and Torres, 1994; UN GA doc of 20 January 1993). It consists of a 5.5-metre parabolic reflector mounted on a rotating alt-azimuth base and allows for the possibility to have receivers at its primary or secondary focus. In order to achieve a homogeneous and calibrated survey, the GEM observatory has been designed for observations with the same instrument from sites at different latitudes. Since 1994 GEM has taken data from Owens Valley (California, USA), Villa de Leyva (Colombia), and Canary Islands (Spain). The availability of galactic emission data in this frequency range has important applications in cosmic ray physics and indirectly in cosmology. Information about the energy spectrum of relativistic cosmic ray electrons can be inferred from a measurement of the radio spectrum. This fact follows from the magnetobremsstrahlung nature of the radio emission in the galaxy. Important contributions to cosmology can also be made using galactic emission data to correct for galactic contamination in maps of the microwave cosmic background radiation (CBR). Studies of anisotropies in the CBR are a powerful tool to test cosmological models, specially since its detection by the DMR receivers on board the Cosmic Background Explorer (COBE) satellite. However, the use of CBR data to extract information about the early universe is severely hampered by the presence of galactic contamination (Haubold and Torres, 1994; Torres, 1995).\par
\medskip
\noindent 
{\bf 2.3 Central America: Honduras' Observatory}\par
\smallskip
\noindent
In Central America, the initiative to establish the first astronomical observatory was born in Honduras at the beginning of the last decade of the twentieth century, pursued by a team of scientists which attended the second UN/ESA Workshop, held at Costa Rica and Colombia in 1992 (UN GA doc of 20 January 1993). Following a strategy, based all the time on the regional cooperation of Central American national universities, and at the international level, establishing close contact with astronomers and prestigious astronomical research centers, the first steps for the establishment of this astronomical observatory has been done (UN GA doc of 18 January 1995). Since 1994, an astronomical observatory is operating in Tegucigalpa, at the Universidad Nacional Autonoma de Honduras (UNAH). This academic unit has been equipped with a 42-cm computerized telescope and other facilities and started with a programme for training of researchers and technicians for the Central American Isthmus. Also, several important agreements of cooperation are making progress in order to contribute to the development of basic space science in the region (Haubold and Onuora, 1994). In the short period of time from 1992 to 1997, under the leadership of astronomers from UNAH, an Assembly of Central American Astronomers and Astrophysicists (AAAC) was established which held their meetings annually in Honduras (1993), Costa Rica (1994), El Salvador (1996), and Panama (1998). Simultaneously, Central American Courses in Astronomy and Astrophysics (CURCAA) were organized in Honduras (1995), El Salvador (1996), Guatemala (1997), and Panama (1998). At the XXIIIrd General Assembly of the IAU at Kyoto, Japan, the six Central American countries became jointly a member of IAU. A crucial role for all these developments played the seventh UN/ESA Workshop, held at Tegucigalpa, Honduras, in 1997, with the inauguration of the Suyapa Observatory at UNAH (Haubold, 1999; UN GA doc of 20 January 1998).\par
\medskip
\noindent
{\bf 2.4 Western Asia: Egypt's Kottamia Telescope}\par
\smallskip
\noindent
In order to maintain Egypt's position in the international astronomical community it was considered desirable to make use of recent developments in mirror-making technology to modernize the approximately 35-year old 1.88-metre Kottamia telescope. Supported by one of the major resolutions of the fourth UN/ESA Workshop, which was hosted by the Government of Egypt in 1994 at Cairo, the question of modernizing the telescope was raised with the Egyptian Government. After extensive discussions between the National Research Institute of Astronomy and Geophysics (NRIAG) at Helwan and the Egyptian Government, this project was approved and funded. The modernization of this telescope is especially important in view of the fact that it is the largest telescope in North and Central Africa, as well as in Western Asia. The importance of modernizing this facility, which would support major experimental capabilities for basic space science in the region, can not be underestimated (Haubold and Mikhail, 1995a, b; UN GA doc of 11 August 1994; Hassan, 1998). The refurbished Kottamia telescope saw first light in September 1997.\par
\medskip
\noindent
{\bf 2.5 Africa: African Skies Newsletter}\par
\smallskip
\noindent 
The Working Group on Space Sciences in Africa is an international, non-governmental organization founded by African delegates at the sixth UN/ESA Workshop, held at Bonn, Germany, 1996 (UN GA doc of 13 December 1996a). The scientific scope of the Working Group's activities is defined to encompass (i) astronomy and astrophysics, (ii) solar-terrestrial interaction and its influence on terrestrial climate, (iii) planetary and atmospheric studies, and (iv) the origin of life and exobiology$^8$. The Working Group seeks to promote the development of space sciences in Africa by initiating and coordinating various capacity-building programmes throughout the region. These programmes fall into a broad spectrum ranging from the promotion of basic scientific literacy in the space sciences to the support of international research projects. The Working Group also promotes international cooperation among African space scientists and acts as a forum for the exchange of ideas and information through its publications, outreach programmes, workshops, and scientific meetings. The newsletter African Skies is published by the Working Group on Space Sciences in Africa (1st issue published in May 1997; 2nd issue April 1998; publication of 3rd issue expected for November 1998). This newsletter is jointly produced by astronomers from the Observatoire de Midi-Pyrenees (France) and the South African Astronomical Observatory (South Africa) and is distributed worldwide, but preferentially to scientific institution in the region of Africa, by the United Nations Office for Outer Space Affairs. The publication of such a newsletter for Africa was long overdue according to observations and recommendations made at UN/ESA Workshops since 1991. While, for example, the number of astronomical papers published annually by members of the American Astronomical Society (AAS) is numerically almost equal to the number of members of AAS (in 1990 about 5500 and growing by 5\% each year). Worldwide the number of astronomical papers (14300 in 1985) is not quite keeping up with the number of members of the IAU. Therefore the growth in astronomical papers is simply due to the growth in numbers of astronomers (Abt, 1993). In the case of Africa (with the exception of South Africa and a number of northern African countries), neither an association of astronomers nor appropriate means for pursuing astronomy, not to speak about publications in astronomy, exist. The African Skies newsletter is supposed to be a very first step to improve this situation in African nations$^9$.\par 
\medskip
\noindent
{\bf 2.6 Astrophysics for Physics Courses at Universities in Developing Nations}\par
\smallskip
\noindent
In all UN/ESA Workshops, the questions on how to teach astronomy and ``teaching astronomy ... but at what level?'' naturally arose in working group sessions, frequently discussed in relation to already existing and recommended teaching material (Narlikar, 1990). All Workshops made observations and recommendations on how to introduce the teaching of astronomy as part of the curriculum for graduate and post-graduate studies, depending on the status of development of astronomy in the respective regions (UN GA doc of 17 December 1991; UN GA doc of 20 January 1993; UN GA doc of 26 May 1994; UN GA doc of 11 August 1994; UN GA doc of 14 May 1996; UN GA doc of 20 January 1998; UN GA doc of 13 December 1996a). To meet the spirit of all these observations and recommendations, a number of participants of the Workshops, particularly during the sixth Workshop held in Germany in 1996, have become involved in developing (i) a booklet Developing Astronomy and Space Science Worldwide (Haubold, 1999); (ii) proposed to bring out a Manual on the Role of Small Telescopes in Education and Research (in preparation); and (iii) to work on a basic ``unit'' on essentials of astronomy with a ``kit'' of simple, hands- on materials, which could be adapted by local educators for effective, appropriate local use (Motz and Duveen, 1977; Percy, 1996).

     For practical reasons, project (iii) will initially be limited to introducing Astrophysics for University Physics Courses (Wentzel and Haubold, 1998). This limitation was mainly based on the notion that physics and mathematics curricula are well developed at almost all universities, while this has not been accomplished at the public school and college level, respectively, in a large number of countries. The booklet (Wentzel and Haubold, 1998) presents an array of astrophysical problems, any one or a few of which can be selected and used within existing physics courses on elementary mechanics, or on heat and radiation, kinetic theory, electrical currents, and in some more advanced courses. These astrophysics problems are designed to be an interesting and challenging extension of existing physics courses, to test the student's understanding of physics by testing it in new realms, and to stretch the student's imagination. A brief tutorial on the astrophysics is provided with each problem, enough so that the physics professor can present the problem in class. All the problems seek compact algebraic and numerical solutions that can easily be translated into physics. The booklet is exposed to an extensive refereeing process by physics professors at universities around the globe, but particularly in developing nations, to secure that it meets the needs and the available curricula and resources at such universities.\par
\bigskip
\noindent 
{\large\bf 3 LONG-TERM PROJECTS OF THE UN/ESA WORKSHOPS}\par
\medskip
\noindent
Since 1991, projects are being pursued within the framework of the UN/ESA Workshops on a long-term basis which are briefly reviewed in the following, with a view to the forthcoming UNISPACE III Conference$^{10}$.\par
\medskip
\noindent
{\bf 3.1 World Space Observatory}\par
\smallskip
\noindent
Based on deliberations of the UN/ESA Workshops, held in 1995 in Sri Lanka (UN GA doc of 14 May 1996), 1996 in Germany (UN GA doc of 13 December 1996a), and 1997 in Honduras (UN GA doc of 20 January 1998), a recommendation was made to explore the feasibility of the establishment of a World Space Observatory. The basic
idea behind the WSO is that general facilities in the windows for astronomical observations which require satellite observatories, are better done through a project with world-wide support, participation and contribution, than with specific projects defined in a more confined limited national configuration. The reasons for this are various (Nature, 1989, 1990, 1994; Wamsteker, 1998):\par
     (i) The needs are essentially similar in most countries, while specific study areas tend to show regional trends, of equivalent scientific value.\par
     (ii) The needs for the stimulation of intellectual capabilities in developing countries can not be supported in their national environment alone with any other possible astronomical facilities (e.g. ground-based or otherwise), at economically viable costs.\par
     (iii) The continued need for studies bearing on the relevance of our place in the Universe requires continued support and can not be driven by addressing the currently popular questions with high prestige projects only.\par
     (iv) A large community of astronomers and astrophysicists will continue to demand support for their science and an extensive interruption of this support over a period of more than a generation can have very dramatic effects on the evolution of knowledge, which is an essential part of our cultural environment in the next millennium.

     The WSO concept could in the long run include space observatories for different wavelengths domains including the X-rays and Gamma-rays, even taking over the operations of projects, launched by major agencies with funding for a limited operations duration. WSO's should not be conceived as technology development projects for the industrialized countries, but as low cost projects where the main emphasis is on the required sensitivity and the stability of operations. Since many aspects of the necessary observatories would possibly not involve the development of the most advanced technologies, but rather rely on well established technologies (as are communications satellites) these projects can be developed in a more cost effective manner than the science projects normally taken on by the major space agencies.

     The current momentum in time is especially suited to the initiation of such a concept for the following reasons:\par
     (i) The concentration of facilities in astronomy to a limited number of high quality facilities is an unstoppable trend.\par
     (ii) A mechanism for indigenous development of science is a necessary prerequisite for the developing world.\par
     (iii) The technology available for communications is sufficiently developed so that the concept can be implemented without disastrous economic burdens for all parties involved.\par
     (iv) Technology for a spacecraft required for such an observatory is today mature technology.\par
     (v) Application of new high technology developed detectors could be an intricate part of such a project without creating a total dependence on new technologies.\par
     (vi) The chance to develop in this context local capabilities for direct and essentially local participation for all countries, presents an enormously attractive possibility to stimulate participation in all levels of society in the exploration of the Universe, especially if it is combined with an assertive public outreach programme.\par
     (vii) A scientific community which has shown to be extremely vigorous, appears to be left without observational opportunities.

     The concept of the WSO has been prepared for presentation to the United Nations Conference on the Exploration and Peaceful Uses of Outer Space (UNISPACE III) (UN GA doc of 20 January 1998; Wamsteker, 1998).\par
\medskip
\noindent
{\bf 3.2 Egypt's Mars Drill Project}\par 
\noindent
During the fourth UN/ESA Workshop, held at Cairo, Egypt, in 1994, the possible participation of Egypt in a future Mars rover mission was discussed (UN GA doc of 3 December 1996b). One concept suggested was that Egypt participate in such a mission through involvement in the design, building, and testing of a drill for obtaining
sub-surface samples (Haubold, 1999).

     The Planetary Society (TPS), a major sponsor of the UN/ESA Workshops, followed-up these discussions. TPS informed the Space Research Institute (IKI) of the Russian Academy of Sciences about this project, and they, in turn, formally invited the Egyptian Ministry of Scientific Research to study the concept for potential use on the Russian Mars 2001 Mission. Of the many important scientific objectives of the Marsokhod mission, among the most interesting, is the analysis of sub-surface samples. Inclusion of some sort of drilling mechanism in the payload of such a mission would assist scientists in the investigation of volatile organic materials and mineralogy.

     Twenty years ago, the arm on the Viking Mars lander was able to obtain samples from depths up to 10-cm. Today, a drill with the capability of boring at least an order of magnitude deeper (more than one metre) would be essential to further research and investigation.

     Egypt has expertise in drill development. Years ago, as part of the archeological exploration of the Pyramids, a sophisticated drilling system was developed to drill into and deploy a camera into a sub-surface chamber without allowing air into the chamber. The drill perforated the limestone to a depth of 2 metre without the use of lubricants or cooling fluids that might have contaminated the pit's environment, and successfully collected six samples (El-Baz, 1997).

     This experience as well as more common terrestrial applications suggest that the necessary technological basis for a drill development can be brought together. In the proposed application for the Mars 2001 Mission, a study team of Egyptian scientists, collaborating with American, Russian, and European scientists, is now pursuing the project.\par
\medskip
\noindent
{\bf 3.3 Network of Oriental Robotic Telescopes}\par
\smallskip
\noindent 
The Network of Oriental Robotic Telescopes (NORT) will mainly deal with variable astronomical objects such as red giants, planetary nebulae, and post-novae, to stimulate asteroseismology of long-period variables and to contribute to progress in the understanding of these objects. Examples of scientific objectives for robotic telescopes are widely appreciated (Filippenko, 1992). As various characteristic times of variation have to be searched for, continuous monitoring of selected typical objects is required. This requirement can only be satisfied through having telescopes at the "best" sites, i.e. semi-desert countries, around the world. 

     A number of such countries are located along the tropic of Cancer, from Morocco to the Chinese deserts. Although a number of these oriental countries had great astronomers in the past, few are now actively participating in astrophysical research. However, their universities, their sites (high mountains in semi- desert climate) and their desire for development can provide strong support for progress in this project (Querci and Querci, 1998).\par

     In the NORT project, it is proposed\par
     (i) to collaborate in astronomy and astrophysics education;\par
     (ii) to help in the development of research laboratories and student observatories within the university context;\par
     (iii) to promote training of engineers and technicians in French observatories such as Haute-Provence Observatory (OHP), Midi-Pyrenees Observatory (OMP); and\par
     (iv) to collaborate in setting up the network and in the scientific choice of the objects to be observed.

     All the equipment will be fully robotic. Each day, all the collected observations will be transmitted directly to all the universities and members of the network via the Internet and/or ARABSAT. The data reduction and interpretation could be done in joint efforts, thus further promoting shared scientific and technical progress.\par
\medskip
\noindent
{\bf 3.4 Southern African Large Telescope}\par
\smallskip
\noindent
As a result of the work of the third UN/ESA Workshop held in Nigeria in 1992, a proposal was drawn up for an Inter-African Astronomical Observatory and Science Park on the Gamsberg in Namibia. Because of its unique geographic location, southern Africa can make an immense contribution to astronomy. Observation of certain time-critical phenomena and 24-hour coverage can be ensured only through astronomical observatories in continents (excluding Antarctica) south of the equator. The Gamsberg has been identified as one of the most suitable sites for an observatory in southern Africa. It is a table mountain 120 kilometres south-west of Windhoek above the Namib desert at an altitude of 2350 metres above sea level. It experiences a large number of cloudless nights, a dark sky, excellent atmospheric transparency and low humidity (Birkle et al,. 1976). The development of an astronomical facility on the Gamsberg can only be achieved, however, through broad international collaboration. After pursuing this project over the period of time from 1992 to 1996, a conclusion was drawn that at this point of time financial and in-kind support is not sufficient to implement this project successfully (Elsaesser, 1996).

     Also part of the deliberations of the UN/ESA Workshop in Nigeria was the discussion of the Southern African Large Telescope (SALT) project which was being developed by South African Astronomical observatories since 1989 (Stobie et al., 1993; Warner 1995). South African astronomers plan to build a 10-m class telescope for optical and infrared astronomy. This telescope will be based on the revolutionary Hobby-Eberly Telescope (HET) nearing completion at McDonald Observatory. Because HET is a specialist design optimized for spectroscopic survey work and imaging over a limited field, it has been built at a fraction of the cost of a conventional 8-m class telescope. SALT will be built near Sutherland, at the observing station of the South African Astronomical Observatory (SAAO). South Africa cannot fund this telescope alone and is seeking international support for its implementation. On the 1st June 1998 the South African Cabinet approved the building of SALT (Stobie ,1997). Since 1995, the United Nations supported SALT by disseminating information on the project to 111 universities in Africa through the Association of African Universities in Ghana.\par
\bigskip
\noindent
{\large\bf 4 CONTINUATION OF THE UN/ESA INITIATIVE INTO THE NEXT MILLENNIUM:
THE SHORT-TERM PERSPECTIVE}\par
\medskip
\noindent
{\bf 4.1 Workshops}\par
\smallskip
\noindent
The next UN/ESA Workshop has been scheduled to be held in March 1999 at the Al al-Bayt University, Mafraq, Hashemite Kingdom of Jordan, inaugurating an astronomical facility at this university$^{11}$.\par
\clearpage
\noindent
{\bf 4.2 Astronomical Telescope Projects}\par
\smallskip
\noindent
Following the example of the establishment of an astronomical telescope facility in Sri Lanka, the science faculty of the Universidad Nacional de Asuncion at Asuncion, Paraguay, is currently making efforts to establish a similar facility for educational purposes for students in physics and astronomy, engineering, meteorology, and geography at this university (Troche-Boggino, 1998). Under the leadership of astronomers and the Government of Japan, a committee was established that explores the possibility of accommodating an astronomical telescope, that might be donated by the Government of Japan to the Government of Paraguay through the Japanese Cultural Grant Aid Programme (Kittamura, 1992). The Facultad Politecnica of the Universidad Nacional de Asuncion will contribute the construction of the observatory building, office space, and an auditorium for the Centre for Astronomy at Paraguay, as this project has been dubbed by the Paraguayan authorities. 

     In creating this project, scientists in Paraguay proudly refer to an early history of astronomy in the country. Buennaventura Suarez was the pioneer astronomer in Paraguay who built a sort of astronomical observatory in the 18th century. He was born in Santa Fe in 1678 and had studied at Cordoba, both cities now located in Argentina. He did most of his work in San Cosme y Damian, one of thirty Jesuit communities for Guarani Indians in the Great Province of Paraguay, until his death in 1750. With the help of local artisans, Suarez built various astronomical instruments, including some Kepler-type refractors with lenses polished from local crystalline rocks, sundials, a quadrant with degrees divided into minutes, and a pendulum clock divided into minutes and seconds. All of Suarez's instruments were lost, with the exception of a sundial at San Cosme y Damian, which now serves as a lonely testament to this exceptional man and early history of astronomy at Paraguay.

     The United Nations is in close contact with astronomers and the Governments of Paraguay and Japan, respectively, to bring the establishment of the Centre for Astronomy at Asuncion, Paraguay, to a fruitful conclusion. Similar projects are being in preparation in the Philippines (optical telescope) (Philippines 1998) and in Egypt (radio telescope) (Mosalam Shaltout, 1998) which have been pushed forward through recent UN/ESA Workshops.\par
\medskip
\noindent
{\bf 4.3 Regional Centres for Space Science and Technology Education}\par
\smallskip
\noindent
A major prerequisite to successful space science and technology applications, in developing countries, is the development of human resources within each region. In recognition of such a need, the UN General Assembly, in its resolution 45/72 of 11 December 1990, endorsed the recommendation that the\par
     ... United Nations should lead, with the active support of its specialized agencies and other international organizations, an international effort to establish regional centres for space science and technology education in existing national/regional educational institutions in the developing countries.

     The UN has been active in the field of space science, primarily through promoting the participation of scientists from developing countries in front-line research and education and through its complementary initiative aimed at establishing regional Centres for Space Science and Technology Education in developing countries and its series of seven UN/ESA Workshops (Centres, 1996).

     The concept behind these regional Centres is based on the fundamental notion that it is vital for developing nations to educate personnel in space science and in the use of space technology, particularly those applications relevant to remote sensing, satellite meteorology, and satellite communications. The establishment of such Centres, one in Asia and the Pacific (India), one in Latin America and the Caribbean (Brazil and Mexico), and one each in the French-speaking (Morocco) and English-speaking (Nigeria) regions of Africa, are underway.

     The regional Centre for Space Science and Technology Education for Asia and the Pacific (CSSTE-AP) was established in Dehra Dun, India, in 1995. From 1st June to 30 November 1998, the fifth postgraduate course, focusing on basic space science, is being conducted at the Physical Research Laboratory, Navrangpura, Ahmedabad, India$^{12}$. This course justifies the decision of UN and ESA not to organize a Workshop in 1998 which was originally scheduled to be held in Tunisia.\par
\medskip
\noindent
{\bf 4.4 The Third United Nations Conference on the Exploration and Peaceful Uses of Outer Space (UNISPACE III)}\par
\smallskip
\noindent
The forthcoming United Nations Conference on the Exploration and Peaceful Uses of Outer Space, to be held from 19 to 30 July 1999 at Vienna (which was also the site of the two previous Conferences in 1968 and 1982), Austria, will comprehensively assess the achievements of the UN/ESA Workshops$^{10}$ (UN GA doc of 20 January 1998). In conjunction with UNISPACE III, the International Astronomical Union (IAU) will hold a UN/IAU Special Environmental Symposium (``Preserving the Astronomical Sky'') and a UN/IAU/COSPAR Special Workshop on Education which will address scientific topics not uncommon to the UN/ESA Workshops (IAU, 1998).\par
\bigskip
\noindent
{\large\bf 5 ACKNOWLEDGEMENTS}\par
\medskip
\noindent
Different versions of this paper have been presented at the XXIIIrd General Assembly of the IAU at Kyoto, Japan, in August 1997, and at the 32nd Scientific Assembly of COSPAR at Nagoya, Japan, in July 1998.

     Without the generous support of the Governments which agreed to host the Workshops, none of them would have materialized.

     The unconditional cooperation with and support from the ten co-organizing entities (ASA, CNES, DARA, ESA, IAU, ICTP, ISAS, NASA, TPS, UN) of the Workshops on Basic Space Science over the period of time of one decade is appreciated.
 
     The organizers of the individual Workshops had embarked on an almost unthinkable mission and their untiring efforts are greatly acknowledged: S.C. Chakravarty (India), W. Fernandez (Costa Rica), S. Torres (Colombia), L.I. Onuora and P.N. Okeke (Nigeria), J.S. Mikhail (Egypt), H.S. Padmasiri de Alwis (Sri Lanka), and M.C. Pineda de Carias (Honduras).

    The help provided by Barbara and Alexander for typesetting the workshop proceedings and maintaining the un-esa homepage, respectively, are gratefully acknowledged.

     Disclaimer: The views, interpretations, and opinions presented in this paper do not necessarily reflect the position of the United Nations.\par
\bigskip
\noindent
{\large\bf 6 REFERENCES}\par
\medskip
\noindent
Abt, H., 1993. Astronomical publications. In K.C. Leung and I.-S. Nha\par 
(eds.), New Frontiers in Binary Research. Astronomical Society of the\par
Pacific Conference Series 38: 466-469.\\ 
Andersen, J.R., 1998. A sequel to an assessment of astronomy in developing\par 
countries by Percy, J.R. and Batten, A.H.,   1995. Chasing the dream.\par 
Mercury, 24: No.2, 15-18.\\
Bahcall, J.N., 1991. The Decade of Discovery in Astronomy and\par 
Astrophysics. National Academy Press, Washington, D.C.\par 
(http://xxx.lanl.gov/abs/astro-ph/9704255).\\
\clearpage
\noindent
Birkle, K., Elsaesser, H., Neckel, Th., Schnur, G. and Schwarz, B., 1976.\par 
Seeing measurements in Greece, Spain, South West Africa, and Chile.\par 
Astronomy and Astrophysics 46: 397-406.\\
Brown, R. (ed.), 1990. An Educational Initiative in Astronomy. Space\par 
Telescope Science Institute, Baltimore.\\
Centres, 1996. Centres for Space Science and Technology Education:\par 
Education Curricula, United Nations, Vienna, pp. 23; and United\par 
Nations General Assembly Document A/AC.105/687 of 19 December\par 
1997.\\
El-Baz, F., Moores, B. and Petrore, C.E., 1997. Remote sensing at an\par 
archeological site in Egypt. American Scientist 77: 60-66; and\par 
El-Baz, F., 1997. Space age archeology. Scientific American,\par 
August, pp. 40-45.\\
Elsaesser, H. (Max-Planck-Institute for Astronomy, Germany), 1995. Letter\par 
to the German Space Agency (DARA), dated May 16. United Nations,\par 
Vienna.\\
Elsaesser, H. (Max-Planck-Institute for Astronomy, Germany), 1996. Letter\par 
to the United Nations, dated September 23. United Nations, Vienna.\\
Fernandez, W. and Haubold, H.J. (eds.), 1993. Basic Space Science. Earth,\par 
Moon, and Planets, 63: No.2, 93-170.\\
Filippenko, A.V. (ed.), 1992. Robotic Telescopes in the 1990s. Astronomical\par 
Society of the Pacific Conference Series Vol. 34.\\
Gehrels, T., 1984. In C. Wickramasinghe (ed.), Fundamental Studies and\par 
the Future of Science. University College Cardiff Press, Cardiff,\par 
pp. 377-385.\\
Gehrels, T., 1988. On the Glassy Sea: An Astronomer's Journey. American\par 
Institute of Physics, New York.\\
Gehrels, T. (ed.), 1994. Hazards Due to Comets and Asteroids.\par 
The University of Arizona Press, Tucson and London.\\
Hassan, S.M., 1998. Upgrading the 1.9-m Kottamia telescope. African Skies,\par 
Number 2, April, pp. 16-17.\\
Haubold, H.J. and Khanna, R.K. (eds.), 1992. Basic Space Science.\par 
Conference Proceedings of the American Institute of Physics Vol. 245,\par 
American Institute of Physics, New York.\\
Haubold, H.J. and Torres, S. (eds.), 1994. Basic Space Science. Astrophysics\par 
and Space Science, 214: Nos.1-2, 1-260.\\
\clearpage
\noindent
Haubold, H.J. and Onuora, L.I. (eds.), 1994. Basic Space Science.\par 
Conference Proceedings of the American Institute of Physics Vol. 320,\par 
American Institute of Physics, New York.\\
Haubold, H.J. and Mikhail, J.S. (eds.), 1995a. Basic Space Science. Earth,\par 
Moon, and Planets, 70: Nos.1-3, 1-233.\\
Haubold, H.J. and Mikhail, J.S. (eds.), 1995b. Basic Space Science.\par 
Astrophysics and Space Science 228: Nos.1-2, 1-405.\\
Haubold, H.J., Ocampo, A., Torres, S., and Wamsteker, W., 1995. United\par 
Nations/European Space Agency workshops on basic space science.\par 
ESA Bulletin, Number 81, February, 18-21.\\
Haubold, H.J., 1996. Astronomy and planetary exploration and the United\par 
Nations. American Astronomical Society Newsletter, Number 79,\par 
March, 17-18. \\
Haubold, H.J. and Wamsteker, W., 1997. Space Technology, Worldwide\par 
development of astronomy: the story of a decade of UN/ESA workshops\par 
on basic space science. (in print;\par 
http://xxx.lanl.gov/abs/astro-ph/9705169).\\
Haubold, H.J. and Mezger, P.G. (eds.), 1998. Basic Space Science.\par 
Astrophysics and Space Science, (in print, contents of the Proceedings\par 
available at\par 
http://www.seas.columbia.edu/$\sim$ ah297/un-esa/ws1996-proceedings.html).\\
Haubold, H.J. and Remo, J.R., 1998. The issues of space debris and\par 
near-Earth objects at the United Nations. In S.Isobe and T. Hirayama\par 
(eds.), Preserving the Astronomical Windows. Astronomical Society of\par 
the Pacific Conference Series 139: 115-118.\\
Haubold, H.J. (ed.), 1999. Developing Astronomy and Space Science\par 
Worldwide. Kluwer Academic Publishers, Doordrecht, (in print;\par 
contents of the book available at\par 
http://www.seas.columbia.edu/$\sim$ ah297/un-esa/ws1995-proceedings.html).\par
IAU, 1998. IAU statement to the UN at\par 
http://www.seas.columbia.edu/$\sim$ ah297/un-esa/ws1998-iau-statement.html.\\
Kitamura, M. (National Astronomical Observatory, Japan), 1992. Letter to\par 
the United Nations, dated June 15. United Nations, Vienna. The\par 
Government of Japan has supported the establishment of moderate-\par
size research telescopes or planetaria for education in south-Asian\par 
developing nations (Burma, Indonesia, Malaysia, Singapore, and \par
Thailand). These efforts have been facilitated through the cultural\par 
grant-in-aid scheme of the Government of Japan.\\
Kitamura, M. (National Astronomical Observatory, Japan), 1994. Letter\par 
to the United Nations, dated September 12. United Nations, Vienna.\par 
Similar to the donation of an astronomical telescope to the\par 
Government of Sri Lanka, the Government of Japan donated planetaria\par 
to developing countries in the Asia and the Pacific region:\par
          (i) GX-type planetarium (GOTO) to the Youth Culture Centre of Burma\par 
(Myanmar) in 1985;\par
          (ii) E-5 type planetarium (GOTO) to the Institute of Marine Technology\par 
of Burma (Myanmar) in 1988;\par
          (iii) Minolta planetarium to the Space Science Education Centre of Malaysia\par 
in 1988;\par
          (iv) GE-II type planetarium (GOTO) to the Haya Cultural Centre of\par 
Jordan in 1990; and\par
          (v) GS-type planetarium (GOTO) to the Burdowan University of India.\\
Kogure, T., 1997. United Nations and astronomy in developing countries -\par 
impressions at the two related workshops. Astronomical Herald of the\par 
Astronomical Society of Japan, 90: No.6, 277-280.\\
Mosalam Shaltout, M.A., 1998. Abu Simple Radio Telescope Project in the\par 
Upper Egypt. Project Document, January, pp. 10.\\
Motz, L. and Duveen, A., 1977. Essentials of Astronomy. Second Edition,\par 
Columbia University Press, New York.\\
Narlikar, J.V., 1990. Curriculum for the training of astronomers.\par 
In J.M. Pasachoff and J.R. Percy (eds.), The Teaching of Astronomy.\par 
Cambridge University Press, Cambridge. Nature 339(1989)574, 344(1990)188,\par 
371(1994)5\\
Philippines, 1998. The Necessity of a High-Grade Telescope for Education\par 
and Future Research Activities in the Philippines: A Project Proposal\par 
Document, Philippine Atmospheric, Geophysical and Astronomical\par 
Services Administration (PAGASA), Department of Science and\par 
Technology (DOST), April 1998, pp. ii+13.\\
Percy, J.R. (ed.), 1996. Astronomy Education: Current Developments,\par 
Future Coordination. Astronomical Society of the Pacific Conference\par 
Series Vol. 89.\\
Querci, F.R. and Querci, M., 1998. African Skies, Number 2, April,\par 
pp. 18-21.\\
\clearpage
\noindent
Remo, J.R. (ed.), 1997. Near-Earth Objects: The United Nations\par 
International Conference. Annals of the New York Academy of Sciences,\par 
822: 1-632.\\
Smith, D.W. and Haubold, H.J. (eds.), 1992. Planetarium - A Challenge for\par
 Educators. United Nations, New York (reprinted 1993); reprinted\par 
United Nations, Vienna, 1994.\\
Stobie, R.S., Glass, I.S., and Buckley, D.A.H., 1993. Southern African Large\par 
Telescope: A Proposal for Funding. South African Astronomical\par 
Observatory, South Africa.\\
Stobie, R.S., 1997. The Southern African Large Telescope Project -\par 
An Invitation to Participate, 16 May, pp. 25; and Nature 393(1998)403.\\
Torres, S., 1995. Space sciences in Latin America: status and opportunities.\par 
In J.P. Vary and G. Violini (eds.), Ciencia y Tecnologia Para America\par 
Central: Plenes y Estrategias. Imprenta Criterio, pp. 177-181.\\
Troche-Boggino, A.E., 1998. Centre for astronomy for Paraguay: a quest for\par 
a moderate-sized telescope.\par 
http://www.seas.columbia.edu/$\sim$ ah297/un-esa/ws1992-centre.html.\\
United Nations General Assembly Document A/AC.105/489 of\par 
17 September 1991; and Astrophysics and Space Science 193(1992)161.\\
United Nations General Assembly Document A/AC.105/530 of 20 January\par 
1993; and Astrophysics and Space Science 204(1993)163-164.\\
United Nations General Assembly Document A/AC.105/560 of 26 May 1994;\par 
and COSPAR Information Bulletin, Number 129, April 1994, pp. 21-22.\\
United Nations General Assembly Document A/AC.105/580 of 11 August\par 
1994; and Earth Space Review 4(1995), Number 2, pp. 28-30.\\
United Nations General Assembly Document A/AC.105/592/Add.1 of\par 
18 January 1995.\\
United Nations General Assembly Document A/AC.105/640 of 14 May 1996;\par 
and COSPAR Information Bulletin, Number 136, August 1996, pp. 8-11.\\
United Nations General Assembly Document A/AC.105/657 of\par 
13 December 1996a; and COSPAR Information Bulletin, Number 138,\par 
April 1997, pp. 21-24.\\
United Nations General Assembly Document A/AC.105/664 of\par 
13 December 1996b.\\
\clearpage
\noindent
United Nations General Assembly Document A/AC.105/682 of\par 
20 January 1998; and COSPAR Information Bulletin, Number 141,\par 
April 1998, pp. 9-10.\\
Vietnam, 1995. Project Document for Building a Cultural Monument:\par 
Planetarium, in Vinh City, under Cultural grant-in-aid scheme of the\par 
Japanese Government, submitted through the Government of Vietnam,\par 
October 1995, pp. 14.\\
Wamsteker, W. (European Space Agency), 1994. Letter to the United\par 
Nations, dated September 13. United Nations, Vienna. The computer\par 
systems had been shipped from ESA to the UN, were checked on\par 
performance by the UN computer department, and were picked-up by\par 
the Permanent Missions of the respective countries to be delivered\par 
to the University of Nigeria (Department of Physics and Astronomy,\par 
Nsukka, Nigeria), Cuban Academy of Sciences (Institute for Astronomy\par 
and Geophysics, La Habana, Cuba), Universidad Nacional Mayor de San\par 
Marcos (Seminario de Astronomia y Astrofisica, Lima, Peru),\par 
University of Ghana (Department of Physics, Accra, Ghana),\par 
Institute of Fundamental Studies (Earth and Space Science Division,\par 
Kandy, Sri Lanka), Universidad Nacional Autonoma de Honduras\par 
(Observatorio Astronomico Centroamericano de Suyapa, Tegucigalpa \par
M.D.C., Honduras).\\
Wamsteker, W., 1998. Declaration from the scientific organizing committee\par 
`` UV astrophysics beyond the IUE final archive'' to the participants\par 
of the conference. In W. Wamsteker and R. Gonzales Riestra (eds.),\par 
Ultraviolet Astrophysics Beyond the IUE Final Archive. ESA SP-413,\par 
February, pp. 849-856.\\
Warner, B., 1986. The research potential of small telescopes.\par 
In J.B. Hearnshaw and P.L. Cottrell (eds.), Instrumentation and\par 
Research Programmes for Small Telescopes. D. Reidel Publishing\par 
Company, Dordrecht, pp. 3-16.\\
Warner, B. (ed.), 1995. Astronomy in South Africa.  Astrophysics and Space\par 
Science 230: 1-503.\\
Wentzel, D.G. and Haubold, H.J., 1998. Astrophysics for University Physics\par 
Courses.  United Nations, Vienna, (currently under review).\\
Wickramasinghe, N.C., 1993. Setting Up of an Astronomical Observatory in\par 
Shri Lanka, Project Proposal, February, pp. 7.\par
\bigskip
\noindent
{\large\bf 7 FOOTENOTES}\par
\noindent
1 Subsequently, these Workshops were co-organized by the Austrian Space Agency (ASA), French Space Agency      (CNES), German Space Agency (DARA), European Space Agency (ESA), International Astronomical Union (IAU),    International Centre for Theoretical Physics Trieste (ICTP), Institute of Space and Astronautical Science of Japan   (ISAS), National Aeronautics and Space Administration of the United States (NASA), The Planetary Society (TPS),   and the United Nations (UN).\\
2 http://www.seas.columbia.edu/$^\sim$ah297/un-esa/index.html\\
3 http://www.naresa.ac.lk/instit/accmt.html\\
4 http://www.unah.hondunet.net/\\
5 http://www.ips-planetarium.org/\\
6 http://www.seas.columbia.edu/$^\sim$ah297/un-esa/planetarium.html\\
7 http://www.seas.columbia.edu/$^\sim$ah297/un-esa/neo.html\\
8 http://da.saao.ac.za/$^\sim$wgssa/\\
9 A similar newsletter (Spanish, English) does exist for the region of Latin America and the Caribbean: Astronomia Latino Americana (ALA). This regional newsletter in the field of astronomy is being edited by astronomers from the Universidad de Guanajuato and INAOE in Mexico and distributed and available electronically (http://www.astro.ugto.mx/$^\sim$ala/). The publication of an astronomical newsletter (Arabic, English) for the region of Western Asia is in preparation by astronomers in Jordan.\\
10 http://www.un.or.at/OOSA/index.html\\
11 http://www.seas.columbia.edu/$^\sim$ ah297/un-esa/ws1998-jordan$_-$astronomy.html\\
12 http://www.seas.columbia.edu/$^\sim$ ah297/un-esa/regional$_-$centres.html\\
\end{document}